\documentclass[%
 reprint,
nofootinbib,
 amsmath,amssymb,
 aps,
]{revtex4-2}

\usepackage{graphicx}% Include figure files
\usepackage{dcolumn}% Align table columns on decimal point
\usepackage{bm}% bold math
\usepackage{appendix}
\begin{document}

\title{Impact of quark quasiparticles on transport coefficients in hot QCD}% Force line breaks with \\

\author{Valeriya Mykhaylova}
\affiliation{%
Institute of Theoretical Physics, University of Wroc{\l}aw, PL-50204 Wroc{\l}aw, Poland
}
\author{Chihiro Sasaki}
\affiliation{%
Institute of Theoretical Physics, University of Wroc{\l}aw, PL-50204 Wroc{\l}aw, Poland
}

%\collaboration{Collaboration}%\noaffiliation

\date{\today}

\begin{abstract}
We study the bulk and shear viscosity and the electrical conductivity in a quasiparticle approach to Yang-Mills theory and QCD with light and strange quarks to assess the dynamical role of quarks in transport properties at finite temperature. 
The interactions with a hot medium are embodied in effective masses of the constituents through a temperature-dependent running coupling extracted from the lattice QCD thermodynamics. 
In Yang-Mills theory, the bulk viscosity to entropy density ratio exhibits a non-monotonous structure around the phase transition temperature. 
In QCD, this is totally dissolved because of a substantial contribution from quark quasiparticles.
The bulk to shear viscosity ratio near the phase transition behaves consistently to the scaling with the speed of sound derived in the AdS/CFT approach, whereas at high temperature it obeys the same parametric dependence as in perturbation theory.
Thus, the employed quasiparticle model is adequate to capture the transport properties in the weak and strong coupling regimes of the theory.
This feature is not altered by including dynamical quarks which, however, retards the system from restoring conformal invariance.
We also examine the individual flavor contributions to the electrical conductivity and show that the obtained behavior agrees qualitatively well with the recent results of lattice simulations and with a class of phenomenological approaches. 

\end{abstract}

\maketitle

%%%%%%%%%%%%%%%%%%%%%%%%%%%%%%%%%%%%%%%%%%%%%%%%%%%%%%%%%%%%%%%%%%%%%%%%%%%%%%%%%%%%%%%%%%%
\section{Introduction\label{Sec:Intro}}
%%%%%%%%%%%%%%%%%%%%%%%%%%%%%%%%%%%%%%%%%%%%%%%%%%%%%%%%%%%%%%%%%%%%%%%%%%%%%%%%%%%%%%%%%%%
Two decades of intensive theoretical explorations of the flow observables in  ideal~\cite{Teaney:2000cw,Huovinen:2001cy,Hirano:2002ds,Broniowski:2008vp,Schenke:2010nt} and viscous~\cite{Romatschke:2007mq,Song:2007fn,Dusling:2007gi,Bozek:2009dw,Schenke:2010rr,Bozek:2012qs,Ryu:2015vwa,Du:2019obx} hydrodynamics 
have successfully delineated the quark-gluon plasma (QGP) created at RHIC and LHC as a strongly-coupled fluid. Thus, its transport properties characterized by the corresponding transport parameters are of particular importance in the evolution of deconfined QCD matter.

The bulk viscosity $\zeta$ indicates the energy dissipation during the expansion of a medium. 
It vanishes in non-interacting systems of massless particles, thus measures the fate of conformal invariance in strongly interacting theories.
The dimensionless ratio of the bulk viscosity to entropy density $\zeta/s$ and the specific shear viscosity $\eta/s$ are the major input in the hydrodynamic equations\mbox{~\cite{LandauLifshitz,Muronga:2006zw,Florkowski:2010zz,Florkowski:2015lra}}, and they reflect a deviation of the medium from local thermodynamic equilibrium.

The bulk viscosity of strongly-interacting matter has been evaluated in various frameworks, e.g., the kinetic quasiparticle models~\cite{Paech:2006st,Torrieri:2007fb,Sasaki:2008fg,Bluhm:2009ef,Chakraborty:2010fr,Plumari:2011mk,Berrehrah:2016vzw,Mitra:2018akk,Soloveva:2019xph}, the parton-hadron string dynamics~\cite{Ozvenchuk:2012kh}, the Nambu-Johna-Lasinio model~\cite{Sasaki:2008um,Son:2004iv,Marty:2013ita,Ghosh:2015mda,Deb:2016myz}, the Polyakov-Quark-Meson (PQM) model~\cite{Singha:2017jmq}, the Green-Kubo formalism~\cite{Kharzeev:2007wb,Karsch:2007jc,Harutyunyan:2017ieu,Czajka:2017bod}, the Chapman-Enskog method~\cite{Mitra:2017sjo}, and the holographic QCD approach~\cite{Gubser:2008yx,Li:2014dsa,Heshmatian:2018wlv}. 
The dynamic criticality of the bulk viscosity has been discussed as a probe of a QCD critical point~\cite{Son:2004iv,Martinez:2019bsn}.

It has been shown in a quasiparticle model~\cite{Bluhm:2009ef} that near the phase transition, the bulk to shear viscosity ratio of a gluon plasma decreases as predicted in the AdS/CFT approach,
whereas at high temperature as in perturbation theory.
This has been further confirmed in models based on the Gribov-Zwanziger quantization~\cite{Florkowski:2015rua,Begun:2016lgx,Jaiswal:2020qmj}.
The approach based on quasiparticle excitations is thus capable to describe a dynamical link between the strong and weak coupling regions of Yang-Mills thermodynamics. 
The same framework has been recently applied to compute the specific shear viscosity at finite temperature, and the role of dynamical quark quasiparticles has been assessed~\cite{Mykhaylova:2019wci}.

The other important parameter is the electrical conductivity $\sigma$. It characterizes the linear response of a system to an external electric field that generates an electrically charged current in the medium.
Thus, it is of relevance in non-central heavy-ion collisions, where strong electric and magnetic fields are expected to emerge~\mbox{\cite{Hirono:2012rt,Tuchin:2013ie,McLerran:2013hla,Gursoy:2014aka}}. 
It has been shown that $\sigma$ quantifies the diffusion of a magnetic field in the medium~\cite{Baym:1997gq} and the soft dilepton emission \cite{Moore:2006qn}, as well as the photon production rate~\cite{FernandezFraile:2005ka,Linnyk:2013wma}.
The electrical conductivity of deconfined matter has been examined in various methods, such as phenomenological quasiparticle models in the relaxation time approximation~\cite{Puglisi:2014sha,Puglisi:2014pda,Thakur:2017hfc,Berrehrah:2016vzw,Soloveva:2019xph,Cassing:2013iz,Singha:2017jmq}, the Green-Kubo formalism~\cite{Puglisi:2014sha,Puglisi:2014pda,Greif:2014oia}, the Chapman-Enskog method~\cite{Mitra:2017sjo,Mitra:2016zdw}, and the Color String Percolation approach~\cite{Sahoo:2018dxn}.

The precise determination of the transport parameters as functions of temperature and chemical potential, as well as their incorporation to the fluid dynamical simulations, is one of the main steps towards understanding the non-trivial evolution of strongly-interacting matter.
This requires comprehension of the dynamical role of light and strange quark quasiparticles on the transport properties.
In particular, their individual contributions to scattering cross sections and the thermodynamically consistent formulation of the transport coefficients are of major importance.

In this paper, we utilize the quasiparticle model (QPM) developed in~\cite{Mykhaylova:2019wci} to study the specific bulk viscosity and the bulk to shear viscosity ratio as well as the electrical conductivity of Yang-Mills and QCD matter. 
We aim at assessing the role of dynamical quark quasiparticles in transport properties in the strong and weak coupling domains of QCD.
In Sec.~\ref{Sec:QPM}, the QPM is briefly outlined and the speed of sound in the two theories is presented with a close comparison to other approaches.
\mbox{In Sec.~\ref{Sec:BulkVisc}} and Sec.~\ref{Sec:Elcond}, the transport coefficients derived in the kinetic theory under the relaxation time approximation~\cite{Hosoya:1983xm,Sasaki:2008fg,Bluhm:2009ef,Chakraborty:2010fr,Thakur:2017hfc} are studied with special emphasis on their flavor dependence.
Finally, we give a brief summary of our results and concluding remarks in~Sec.~\ref{Sec:Conclusions}.

%%%%%%%%%%%%%%%%%%%%%%%%%%%%%%%%%%%%%%%%%%%%%%%%%%%%%%%%%%
\section{Quasiparticle Model\label{Sec:QPM}}
%%%%%%%%%%%%%%%%%%%%%%%%%%%%%%%%%%%%%%%%%%%%%%%%%%%%%%%%%%%%%%%%%%%%%%%%%%%%%%%%%%%%%%%%%%%

We employ the well-established quasiparticle model~\cite{Bluhm:2004xn} to study transport properties of the QGP
above but not far from the deconfinement phase transition.
The main building blocks are quasiparticle excitations with effective masses depending on temperature and chemical potential, and
the QGP is described as a dynamical fluid composed of the quasiparticles.
In this paper, we restrict ourselves to studying the QGP at finite temperature and vanishing chemical potential.

In thermal equilibrium the quasiparticles are assumed to propagate on-shell with energies $E_i=\sqrt{p^2+m_i^2}$, where $p$ represents a three-momentum
and $m_i$ is the effective mass of a particle species $i$ given by 
\begin{eqnarray}
\label{equ:effmass}
m_i^2=(m_i^0)^2+\Pi_i.
\end{eqnarray}
Here, $m_{i}^0$ is a bare mass of the quasiparticle and $\Pi_i$ is the dynamically generated self-energy. For the QGP composed of weakly-interacting light (degenerate up and down) quarks, strange quarks and gluons, we set the bare masses at $m_l^0=5~\textrm{MeV}$, $m_s^0=95~\textrm{MeV}$ and $m_g^0=0$. The quasiparticles at a given temperature $T$ are characterized by the gauge-independent hard thermal loop (HTL) self-energies $\Pi_i$ with the asymptotic forms~\cite{Pisarski:1989wb,Bluhm:2006yh}
\begin{eqnarray}
\label{e:piq}
\Pi_l(T) & = & 2 \left(m_l^0 \sqrt{\frac{G(T)^2}{6}T^2}+
\frac{G(T)^2}{6}T^2\right) , \\
\label{e:pis}
\Pi_s(T) & = & 2 \left(m_s^0 \sqrt{\frac{G(T)^2}{6} T^2}+
\frac{G(T)^2}{6} T^2\right),\\
\label{e:pig}
\Pi_g(T) & = & \left(3+\frac{N_f}{2}\right)\frac{G(T)^2}{6}T^2.
\end{eqnarray}

\begin{figure}[t]
	\centering
	\includegraphics[width=0.98\linewidth]{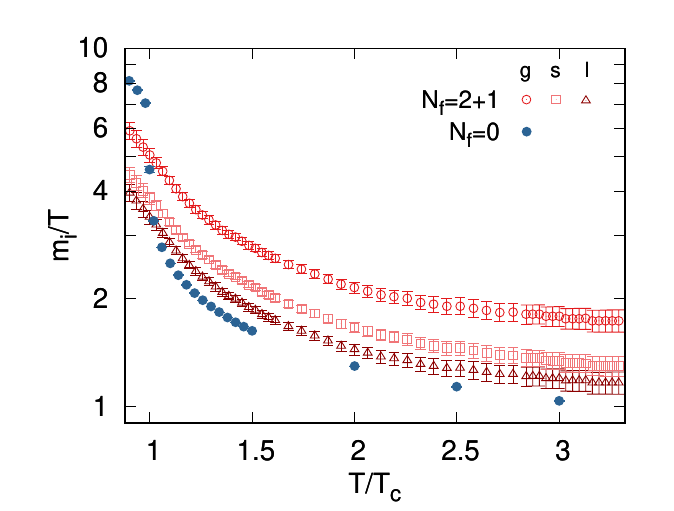}
	\caption{Effective quasiparticle masses scaled with temperature as functions of $T/T_c$. Open bullets represent dynamical masses of gluons (circles), strange (squares) and light (triangles) quarks in QCD with $N_f=2+1$, whereas full circles stand for the mass of gluons in pure Yang-Mills theory. The error bars are due to the uncertainties of the entropy density~$s/T^3$ in the lattice data~\cite{Borsanyi:2013bia,Borsanyi:2012ve}.
We use $T_c = 155$ MeV for $N_f=2+1$ and $T_c = 260$ MeV for $N_f=0$.}
	\label{fig:massestot}
\end{figure}

In order to incorporate non-perturbative features near the QCD phase transition into the model,
we introduce an effective running coupling $G(T)$ which can be extracted from the equation of state calculated in lattice gauge theory.
We utilize the lattice results of the entropy density for the QGP with $2+1$ quark flavors~\cite{Borsanyi:2013bia} and for pure SU(3) Yang-Mills
theory~\cite{Borsanyi:2012ve}. The resultant coupling $G(T)$ captures the non-trivial dynamics near a crossover with $N_f=2+1$
(a first-order phase transition with $N_f=0$), and reproduces the perturbative behavior in the very high-temperature regime~\cite{Mykhaylova:2019wci}.
Figure~\ref{fig:massestot} shows the effective masses calculated via Eq.~(\ref{equ:effmass}) scaled with temperature in pure Yang-Mills theory
and in QCD with $N_f=2+1$.
Their characteristic behavior is a direct consequence of the effective running coupling $G(T)$ extracted according to the aforementioned prescription.
For $N_f=0$, an abrupt change in the effective gluon mass near $T_c$ is responsible for describing a jump in the entropy density at the first-order
phase transition.
For $N_f=2+1$, the temperature profile of $G(T)$ becomes much milder and smoother at any temperature. The hierarchy in the effective masses are
in accordance with the flavor indices and bare-mass dependence in \mbox{Eqs.~(\ref{equ:effmass})-(\ref{e:pig})~\cite{Mykhaylova:2019wci}}.
All the scaled effective masses vanish logarithmically at high temperatures, matching the perturbative expectation for particles with thermal momenta
$p\sim T$~\cite{LeBellac}.

%%%%%%%%%%%%%%%%%%%%%%%%%%%%%%%%%%%%%%%%%%%%%%%%%%%%%%%%%%%%%%%%%%%%%%
\begin{figure*}[t]
	\centering
	\includegraphics[width=0.48\linewidth]{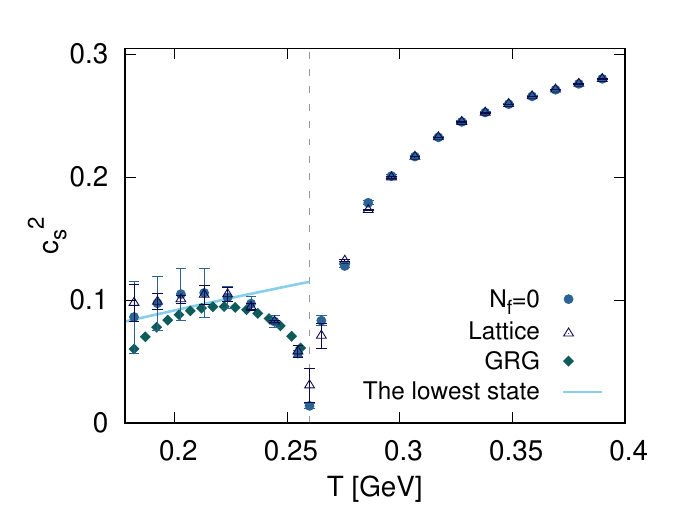}
	\includegraphics[width=0.48\linewidth]{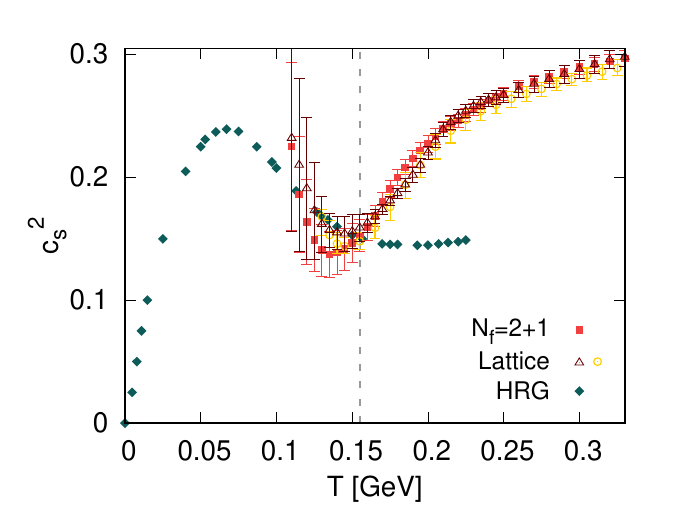}
	\caption{Speed of sound squared as a function of temperature. Left: The result for pure Yang-Mills theory obtained in the quasiparticle model (circles) is compared with $c_s^2$ deduced from the lattice data in Ref.~\cite{Borsanyi:2012ve} (triangles),
from the glueball resonance gas with the Hagedorn spectrum~\cite{Borsanyi:2012ve, Meyer:2009tq} (diamonds),
and  from the ideal gas of the lowest glueball ~\cite{Castorina:2009de, Khuntia:2016ikm} (solid line).
Right:~The same quantity but for $N_f=2+1$ (squares), in comparison to the corresponding result of the lattice QCD simulations~\cite{Borsanyi:2013bia} (triangles) and \cite{Bazavov:2014pvz} (circles) and to the hadron resonance gas with the states below $2.5$~GeV~\cite{Castorina:2009de, Khuntia:2016ikm}~(diamonds). The dashed vertical lines indicate the (pseudo)critical temperatures in both theories. }
	\label{fig:cs2}
\end{figure*}
%%%%%%%%%%%%%%%%%%%%%%%%%%%%%%%%%%%%%%%%%%%%%%%%%%%%%%%%%%%%%%%%%%%%%%%

%%%%%%%%%%%%%%%%%%%%%%%%%%%%%%
\subsection{Speed of sound\label{Subsec:cs2}}
%%%%%%%%%%%%%%%%%%%%%%%%%%%%%

In the quasiparticle model, all thermodynamic quantities in thermal equilibrium are expressed as standard phase-space integrals over the distribution functions. The entropy density for $N_f=2+1$ at vanishing chemical potential is given by the sum of contributions from light (as a sum of up and down) quarks $l$, strange quarks~$s$, their anti-particles and gluons $g$ as
\begin{eqnarray}
\label{e:entr0}
s = \sum_{i = l,\bar{l},s,\bar{s}, g} \frac{d_i}{2\pi^2} \int\,
\!\!\! dp \ p^2 \frac{\left( \frac{4}{3}p^2{+}m_i^2 \right)}{E_i T}
f_i^0 \,, \label{entropy}
\end{eqnarray}
where the spin-color degeneracy factor $d_i$ reads explicitly as \mbox{$d_{l,\, \bar{l}}=2N_cN_l=12$} for  \mbox{$N_l=2$} light \mbox{(anti-)quarks}, \mbox{$d_{s,\, \bar{s}}=2N_c=6$} for strange \mbox{(anti-)quarks} and
\mbox{$d_g=2(N_c^2-1)=16$} for gluons; $f_i^0= (\exp(E_i/T)\pm 1)^{-1}$ denotes the standard distribution function for fermions with the upper sign
and for bosons with the lower sign.
In pure Yang-Mills theory, the thermodynamics of a gluon plasma is obtained by setting $N_f=0$, i.e., $d_{l,\, \bar{l},\, s,\, \bar{s}}=0$.

As we will see in the next section, the speed of sound is one of the essential building blocks of the bulk viscosity to measure a deviation
from the conformal limit.
The speed of sound squared is obtained once the entropy density is calculated as a function of temperature via
\begin{eqnarray}
c_s^2=\frac{\partial P}{\partial \epsilon}= \frac{s}{T} \left(\frac{\partial s}{\partial T}\right)^{-1} \label{e:cs2},
\end{eqnarray}
where  $P$ denotes the pressure, $\epsilon$ the energy density and $s$ the entropy density calculated in Eq.~(\ref{e:entr0}).
The results are presented in Fig.~\ref{fig:cs2} for pure Yang-Mills theory (left) and for QCD with $N_f=2+1$ (right).

In the left panel, one readily finds that the speed of sound squared of the gluon plasma in the QPM is in excellent agreement, both in confined
and deconfined phases, with the results deduced from the lattice data for the pressure and energy density in pure Yang-Mills theory~\cite{Borsanyi:2012ve}.
This arises from the effective running coupling $G(T)$ defined with the entropy density in the same lattice setup.

It is also instructive to compare the QPM result with the model for a glueball resonance gas (GRG). This can be done along with the parametric form
for the entropy density suggested in~\cite{Borsanyi:2012ve},
\begin{eqnarray}
\frac{s_{\textrm{conf}} (T)}{ T^3 }=\Big( -0.2 \frac{T}{T_c} - 0.134\, F(T)\Big),
\label{e:entropyHRG}
\end{eqnarray}
with $F(T)=\log \Big[1.024 -  \frac{T}{T_c} \Big]$.
This includes the contribution from the GRG beyond the two-particle threshold, i.e., including the Hagedorn density of states~\cite{Meyer:2009tq}.
%with masses above $2 m_0$, where $m_0$ is a mass of the lightest glueball~\cite{Meyer:2009tq}. Here we set $m_0=2$ GeV.
The resultant $c_s^2$ is found easily as
\begin{equation}
c_s^2 = \frac{(1.024\, T_c-T)\Big( 0.2 \frac{T}{T_c} + 0.134\, F(T)\Big)}{(0.412\, T_c-0.402\, T) F(T) + T\Big(0.686-0.8 \frac{T}{T_c}\Big)},\hspace{-0.11cm}
\end{equation}
which well captures the behavior near $T_c$ as seen in the figure.

As a useful reference, one takes a simple model for an ideal bosonic gas including only the lowest glueball.
The speed of sound squared is calculated analytically as in the form~\cite{Castorina:2009de, Khuntia:2016ikm}
 \begin{eqnarray}
 c_s^2=\Big(3+ \frac{m_0^2 K_2 (m_0/T)}{4 T^2 K_2(m_0/T)+ m_0T K_1 (m_0/T)}   \Big)^{-1},
\label{cs2bessel}
 \end{eqnarray}
where $K_{1,2}$ are the modified Bessel functions of the second kind. The parameter $m_0$ denotes the glueball mass and we take \mbox{$m_0=2$~GeV}.
The comparison with the GRG and the QPM approaches as well as the lattice result clearly illustrates that it is insufficient
to describe the thermodynamics near $T_c$ with the lowest state only, and~Eq.~(\ref{cs2bessel}) fails even qualitatively although it describes
the~$c_s^2$ better at lower temperature.

In Fig.~\ref{fig:cs2} (right), we present the $c_s^2$ of the QGP with $N_f=2+1$ quark flavors in the QPM and hadron resonance gas (HRG) model as well as those in lattice QCD.
Within the errors, the overall behavior of the QPM result is fairly consistent with lattice QCD~\cite{Borsanyi:2013bia, Bazavov:2014pvz}.
The HRG model~\cite{Castorina:2009de} describes the $c_s^2$ rather well near the crossover and an apparent deviation from the lattice data emerges
just above $T_c$, indicating that the hadronic picture of the QCD thermodynamics breaks down.

The $c_s^2$ exhibits a non-monotonicity around the corresponding $T_c$ in the two theories, whereas this behavior is much stronger in pure
Yang-Mills theory, as a consequence of the rapid change with temperature in the entropy density at the first-order phase transition.
At higher temperature, the $c_s^2$ approaches the Stefan-Boltzmann limit $\partial P/\partial \epsilon = 1/3$. We will explore how the system
recovers its conformality depending on the quark flavors in the next section.

\begin{figure*}[htb!]
	\centering
	\includegraphics[width=0.48\linewidth]{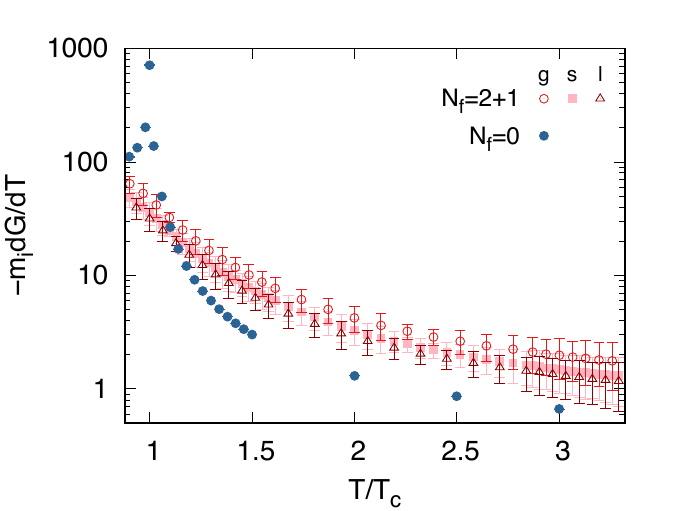}
	\includegraphics[width=0.48\linewidth]{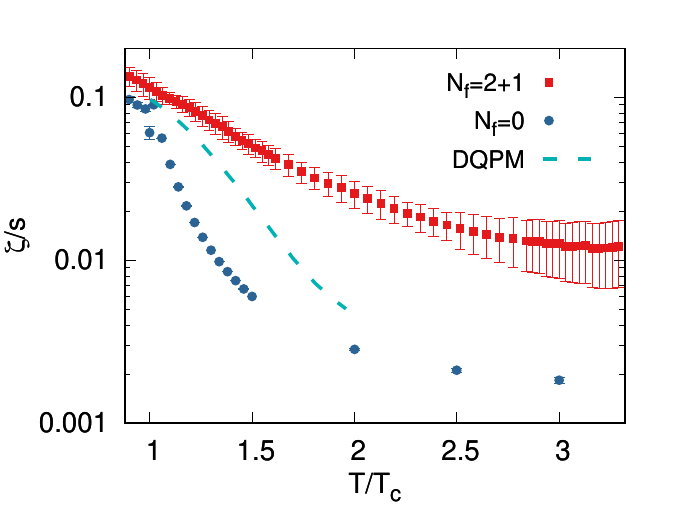}
	\caption{Left: Temperature derivative of the effective coupling multiplied by the masses as a function of $T/T_c$. The result in pure Yang-Mills theory (full circles) is compared to those for each particle species in QCD with $2+1$ flavors: gluons (open circles), strange quarks (squares) and light quarks (triangles). Right: Bulk viscosity to entropy density ratio as a function of~$T/T_c$ for the gluon plasma (full circles) and QGP (full squares). The dashed line corresponds to the dynamical quasiparticle model (DQPM) result for ${N_f=2+1}$~\cite{Soloveva:2019xph}.}	
	\label{fig:CouplingDer+BulkSQPM}
\end{figure*}

We have explicitly demonstrated that the QPM captures the non-perturbative properties of the bulk thermodynamic quantities not only in deconfined phase
but also somewhat below $T_c$ in pure Yang-Mills and full QCD. However, we emphasize that such an agreement with the hadronic picture in confined phase would not be expected for transport coefficients since they carry the details of kinematics with entirely different constituents, i.e., hadrons versus quarks and gluons.
We will therefore restrict our temperature range to the domain of $T/T_c > 0.9$ in the subsequent sections.

%%%%%%%%%%%%%%%%%%%%%%%%%%%%%%%%%%%%%%%%%%%%%%%%%%%%%%%%%%%%%%%%%%%%%%%%%%%%%%%%%%%%%%%%%%%
\section{Bulk Viscosity\label{Sec:BulkVisc}}
%%%%%%%%%%%%%%%%%%%%%%%%%%%%%%%%%%%%%%%%%%%%%%%%%%%%%%%%%%%%%%%%%%%%%%%%%%%%%%%%%%%%%%%%%%%

Assuming that a system deviates from thermal equilibrium only slightly, we can determine the transport parameters via the Boltzmann kinetic equation.
In the relaxation time approximation, the bulk viscosity of a hot matter composed of quark and gluon quasiparticles reads~\cite{Chakraborty:2010fr, Bluhm:2009ef}
\begin{eqnarray}
\zeta=\frac{1}{T}\sum_{i=l,\bar{l}, s, \bar{s}, g}\int \frac{d^3p}{(2 \pi)^3}d_i f_i^0 (1\pm f_i^0) \frac{\tau_i}{E_i^2} \nonumber\\ \times \Big\{\Big(E_i^2-T^2 \frac{\partial \Pi_i(T)}{\partial T^2}\Big)\frac{\partial P}{\partial \epsilon}-\frac{p^2}{3}\Big\}^2, \label{zetaeq}
\end{eqnarray}
where the upper (lower) sign corresponds to Fermi-Dirac (Bose-Einstein) statistics, $\tau_i$ denotes the energy-averaged relaxation time and ${\partial P}/{\partial \epsilon}$ is the speed of sound squared given by Eq.~(\ref{e:cs2}). One can deduce the bulk viscosity of a gluon plasma by setting
$N_f=0$ and $i=g$, with the effective coupling for pure Yang-Mills theory.

The collision term of the Boltzmann equation is parameterized with the relaxation time $\tau$ defined by \mbox{${\tau}^{-1}={n}{\bar{\sigma}}$},
with the equilibrium number density $n$ and the thermal-averaged total cross section $\bar{\sigma}$ for microscopic scattering processes
between the medium constituents.
In~pure Yang-Mills theory, the relaxation time is quantified by elastic gluon-gluon interactions as
${\tau_g}^{-1}={n_g}{\bar{\sigma}_{gg\rightarrow gg}}$.
In QCD with $N_f=2+1$ quark flavors, the QGP as a multi-component medium is characterized by a set of relaxation times.
This is given conveniently in the following matrix form~\cite{Hosoya:1983xm, Mykhaylova:2019wci}
\begin{eqnarray}
\begin{pmatrix}
\tau^{-1}_l \\
\tau^{-1}_{\bar{l}} \\
\tau^{-1}_s \\
\tau^{-1}_{\bar{s}} \\
\tau^{-1}_g
\end{pmatrix}
=
\begin{pmatrix}
\bar{\sigma}_{ll} & \bar{\sigma}_{l\bar{l}} & \bar{\sigma}_{ls} & \bar{\sigma}_{l\bar{s}} & \bar{\sigma}_{lg} \\
\bar{\sigma}_{\bar{l}l} & \bar{\sigma}_{\bar{l}\bar{l}} & \bar{\sigma}_{\bar{l}s} & \bar{\sigma}_{\bar{l}\bar{s}} & \bar{\sigma}_{\bar{l}g} \\
\bar{\sigma}_{sl} &\bar{\sigma}_{s\bar{l}} & \bar{\sigma}_{ss} & \bar{\sigma}_{s\bar{s}} & \bar{\sigma}_{sg} \\
\bar{\sigma}_{\bar{s}l} & \bar{\sigma}_{\bar{s}\bar{l}} & \bar{\sigma}_{\bar{s}s} & \bar{\sigma}_{\bar{s}\bar{s}} & \bar{\sigma}_{\bar{s}g} \\
\bar{\sigma}_{gl} & \bar{\sigma}_{g\bar{l}} & \bar{\sigma}_{gs} & \bar{\sigma}_{g\bar{s}} & \bar{\sigma}_{gg}
\end{pmatrix}
\begin{pmatrix}
n_l \\
n_{\bar{l}} \\
n_s \\
n_{\bar{s}} \\
n_g
\end{pmatrix}\,,
\label{matrix}
\end{eqnarray}
with the equilibrium number density for each species
\mbox{$n_{i} =  \int d^3p/(2\pi)^3 d_i\, f_i^0$},
and $\bar{\sigma}_{ij}$ being the thermal-averaged total cross sections for the two-body elementary scattering processes between the quasiparticles $i$
and $j$. The cross sections are evaluated at tree level with the Feynman propagators of quarks and gluons which carry the effective masses introduced in Eq.~(\ref{equ:effmass}).

The relaxation time approximation is valid in a diluted system, i.e., when the mean free path $\lambda$ is greater than the average interparticle distance $d$,
\mbox{$\lambda \sim \tau \gg d \sim n^{-1/3}$} \cite{Danielewicz:1984ww,Zhuang:1995uf}, where $n=\sum n_i$ is the total particle number density of the system.
In the QPM formulated in~\cite{Mykhaylova:2019wci} we find that at $T_c$ in pure Yang-Mills theory, \mbox{$\tau \sim  0.4$ $\textrm{fm}$} and $d \sim 10^{-4}$ $\textrm{fm}$. 
When temperature reaches $3\ T_c$, the relaxation time remains of the same order, whereas the average distance between the quasiparticles 
decreases to the order of $10^{-6}$ $\textrm{fm}$.
Thus, the condition, $\tau \gg d$, is satisfied in the whole range of temperature considered in this paper.
Similar numbers satisfying the condition are also obtained in QCD with $N_f=2+1$~\cite{Mykhaylova:2019wci}.

Our major assumption in this study is that all the transport parameters for a given particle species carry a common relaxation time. Each parameter is
characterized by a particular dissipative phenomenon formed in the viscous fluid, thus, the corresponding relaxation times are in general different.
The shear viscosity emerges because of the longitudinal fluid motion, hence it is sensitive to the changes of the transverse momentum density, which are
carried on the microscopic level by the elastic $2 \rightarrow 2$ scattering processes included in Eq.~(\ref{matrix}).
The bulk viscosity, on the other hand, characterizes the diffusion of the particles during a uniform expansion of the medium, therefore its
relaxation time essentially depends on the inelastic collisions changing the number density of the excitations~\cite{Jeon:1995zm}.
In this context, within a scalar field theory~\cite{Czajka:2017bod}, response functions of the energy-momentum tensor have been carefully examined
to derive the shear and bulk relaxation times.
Further, electrical conductivity measures the transfer of the electric charge, separately from the momentum transfer, resulting in a different
relaxation time from those for the shear and bulk viscosities.
We will not take those complications into account, but we rather aim at clarifying the dynamical role of the quasiquarks in the QGP.
We also note that the relaxation time defined above is independent of momentum since it is introduced as a mean. 
With given scattering amplitudes,
one can evaluate the momentum-dependent relaxation times including inelastic collisions as instructed in~\cite{Chakraborty:2010fr}. To justify that the averaged relaxation time adequately assesses the properties of the transport parameters, we also compute the $\zeta/s$ ratio in pure Yang-Mills theory using the energy-dependent relaxation time $\tau(s)$ in Appendix~\ref{sec:appendix}.

%%%%%%%%%%%%%%%%%%%%%%%%%%%%%%%%%%%%%%%%%%%%%%%%%%%%%%%%%%%%%%%%%%%%%%%
\subsection{Bulk viscosity to entropy density ratio\label{Subsec:BulkToEntropy}}	

The term $-\partial\Pi_i/\partial T^2$ in Eq.~(\ref{zetaeq}) readily generates a temperature derivative of the effective coupling
in the form of $-m_i\,dG/dT$, displayed in Fig.~\ref{fig:CouplingDer+BulkSQPM} (left).
In~pure Yang-Mills theory, the derivative $- m_g\,dG/dT$ exhibits a prominent maximum at the critical temperature $T_c$, whereas in QCD with $N_f=2+1$,
for any type of constituents it varies smoothly and the strong non-monotonicity seen in the $N_f=0$ case disappears.
The resultant bulk viscosity to entropy density ratio $\zeta/s$ is shown in Fig.~\ref{fig:CouplingDer+BulkSQPM} (right), and one finds that
the characteristic features of the quasiparticle masses and their thermal profiles are encoded in the ratio.
The presence of light and strange quasiquarks causes a significant delay of the QGP approaching a non-interacting gas with $\zeta \to 0$ at high temperature.

In Fig.~\ref{fig:CouplingDer+BulkSQPM} (right), the QPM result for the  QGP with $N_f=2+1$ is confronted with that evaluated in the dynamical
quasiparticle model (DQPM)~\cite{Soloveva:2019xph}.
The overall behaviors as functions of $T/T_c$ in the two approaches are similar, whereas the $\zeta/s$ in the DQPM decreases much faster as
temperature increases.
The observed difference can be traced back to the fact that in the DQPM, the quasiparticles carry finite lifetimes, which reproduce the same
lattice equation of state but modify the expressions for the relaxation times. This may explain the gap between the two results
at high temperature~\footnote{
In addition, as the standard prescription, we evaluated the total cross sections in the large angle scattering (LAS) approximation~\cite{Zhuang:1995uf,Sasaki:2008um,Danielewicz:1984ww}, while this is not used in~\cite{Soloveva:2019xph}.
When the LAS approximation is relaxed, the specific shear viscosity $\eta/s$ decreases by a factor of $\sim 2/3$~\cite{Mykhaylova:2019wci}, and appears close to the result of the Bayesian analysis comparing a hydrodynamical model to the experimental data~\cite{Auvinen:2020mpc}.
}. 
\begin{figure*}[t]
		\centering
	\includegraphics[width=0.48\linewidth]{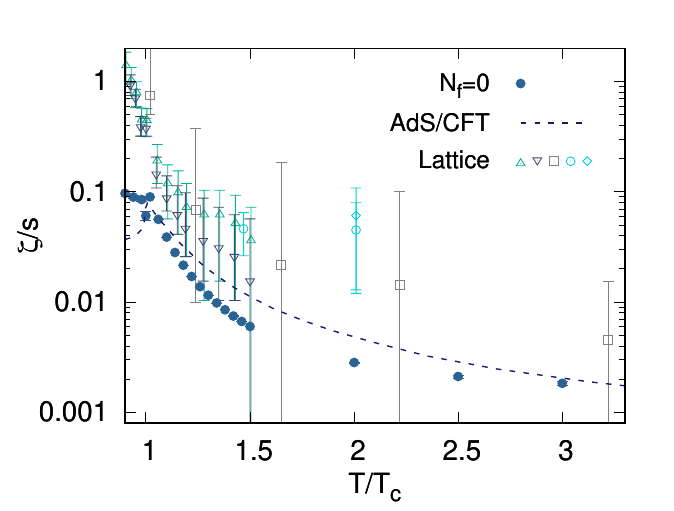}
	\includegraphics[width=0.48\linewidth]{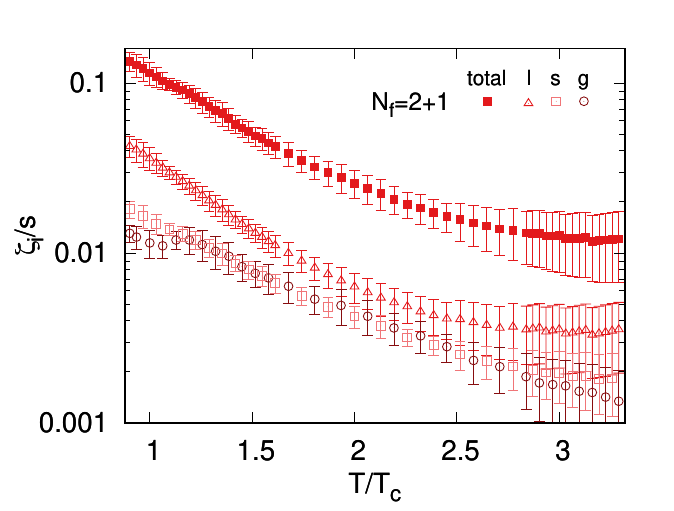}
	\caption{Left: Bulk viscosity to entropy density ratio in pure Yang-Mills theory (full circles). For comparison, we show the corresponding lattice gauge theory results from \cite{Astrakhantsev:2018oue} (open triangles), \cite{Meyer:2007dy} (open squares) and \cite{Sakai:2007cm} (open circles and diamond) as well as the holographic result from \cite{Gubser:2008yx} (dashed line). Right: The same quantity but for QCD with $N_f=2+1$ quark flavors. The total specific bulk viscosity (full squares) is compared to the individual contributions coming from light quarks~(triangles), strange quarks~(open squares) and gluons~(circles). The anti-particle contributions are not included in the ratios $\zeta_{l,\, s,\, g}/s$.}
		\label{fig:ZetaSYMlQCD}
\end{figure*}

In Fig.~\ref{fig:ZetaSYMlQCD} (left), the $\zeta/s$ ratio in pure Yang-Mills theory is compared with the results in other approaches.
The QPM result above $T_c$ is fairly consistent to the collected data sets from lattice gauge theory~\cite{Sakai:2007cm,Meyer:2007dy,Astrakhantsev:2018oue}
as well as to that from an approach based on the gauge-gravity correspondence~\cite{Gubser:2008yx}.
In confined phase, the lattice results show that the ratio continuously decreases as temperature increases toward $T_c$, but this behavior is not
captured either by the holography or the QPM. As emphasized in Sec.~\ref{Sec:QPM}, the QPM is not capable to correctly describe the kinematics of
glueballs, which is essential to evaluate the total cross sections below $T_c$. Therefore, the interpretation of the bulk viscosity should be made
with caution. The observed minimum right below $T_c$ and the result at lower temperature might be the artifacts of the quasiparticle approximation,
and they require further justifications in a more refined approach which resembles confinement.

The right panel of Fig.~\ref{fig:ZetaSYMlQCD} shows the total specific bulk viscosity of the QGP for~$N_f=2+1$ along  with the contributions coming from different quasiparticle species. The $\zeta_{l,\, s,\, g}/s$ ratios are evaluated individually in Eq.~(\ref{zetaeq}) devided by the total entropy density given by Eq.~(\ref{entropy}). The light quarks bring the main impact to the total bulk viscosity of the QGP, while the contributions of strange quarks and gluons are relatively suppressed by their larger effective masses, as in Fig.~\ref{fig:massestot}. We find that the strange quarks and gluons contribute almost equally to the bulk viscosity coefficient via different quantum statistics encoded in the characteristic derivatives of the self-energies. 
The quantitative resemblance between $\zeta_s/s$ and $\zeta_g/s$ comes from the convolution of the degeneracy factors $d_g>d_s$, the relaxation times $\tau_s>\tau_g$, and the effective masses $m_g>m_s$ entering the corresponding energies $E_i$ in Eq.~(\ref{zetaeq}).
This is a clear distinction to the specific shear viscosity, in which the strange-quark component is larger at any temperature than the contribution from gluons~\cite{Mykhaylova:2019wci}. 

%%%%%%%%%%%%%%%%%%%%%%%%%%%%%%%%%%%%%%%%%%%%%%%%%%%%%%%%%%%%%%%%%%%%%%
\begin{figure*}[t]
	\includegraphics[width=0.48\linewidth]{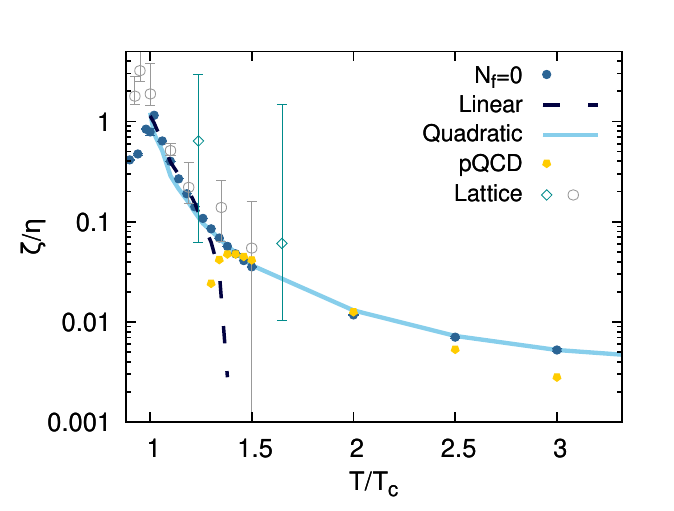}
	\includegraphics[width=0.48\linewidth]{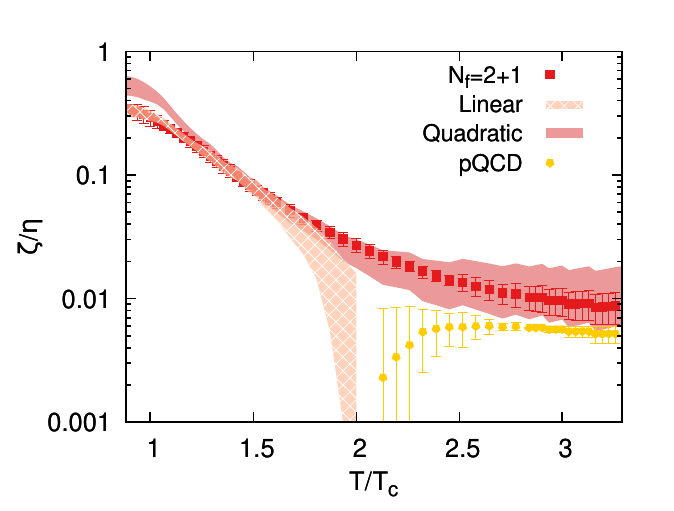}
	\caption{The bulk to shear viscosity ratio as a function of $T/T_c$. The QPM results are compared with the linear and quadratic dependence on the squared speed of sound, Eqs.~(\ref{approx_lin})~and~(\ref{approx_quad}), with the fit parameters
$\alpha=4.5, \beta=-0.3, \gamma=12, \delta=0.002$ for Yang-Mills theory (left)  and $\alpha=2.15, \beta=-0.085, \gamma=14, \delta=0$ for QCD with $N_f=2+1$ (right).
Left: $\zeta/\eta$ in pure Yang-Mills theory~(full circles) parameterized by linear (dashed line) and quadratic (solid line) ansatzes. The results deduced from the lattice data in~\cite{Meyer:2007ic, Meyer:2007dy}~(open diamonds) and \cite{Astrakhantsev:2017nrs, Astrakhantsev:2018oue}~(open circles) as well as those from perturbative QCD~\cite{Arnold:2003zc,Arnold:2006fz} (full pentagons) are shown for comparison.
Right: The same quantity but in $N_f=2+1$ QCD~(full squares), shown along with  linear (checkered band) and quadratic (plain-colored band) parametrizations.}
	\label{fig:BulkShearApprox}
\end{figure*}
%%%%%%%%%%%%%%%%%%%%%%%%%%%%%%%%%%%%%%%%%%%%%%%%%%%%%%%%%%%%%%%%%%%%
%%%%%%%%%%%%%%%%%%%%%%%%%%%%%%%%%%%%%%%%%%%%%%%%%%%%%%%%%%%%%%%%%%%%%%%%
\begin{figure*}[t]
	\includegraphics[width=0.48\linewidth]{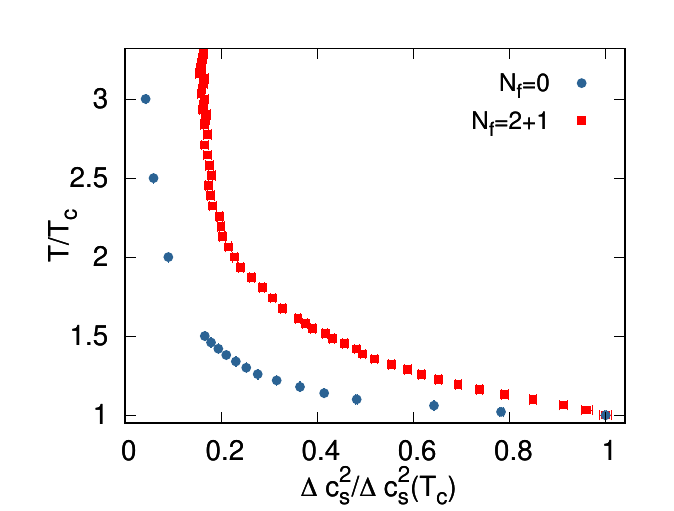}
	\includegraphics[width=0.48\linewidth]{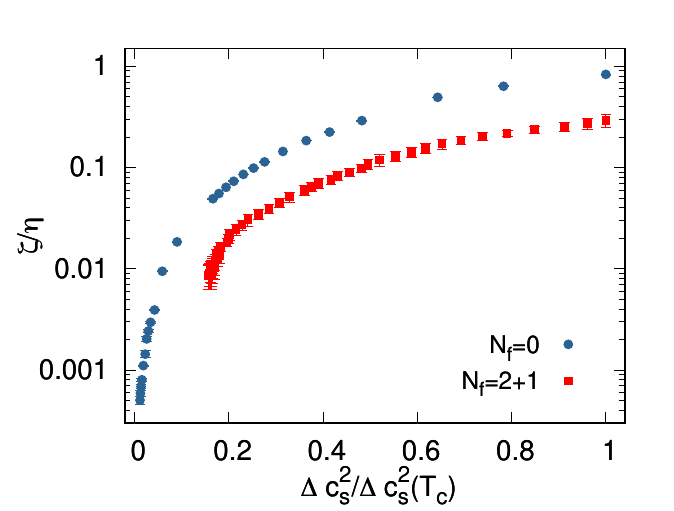}
	\caption{The scaled temperature (left) and the bulk to shear viscosity ratio (right) as functions of the conformality measure, $\Delta c_s^2 = 1/3 - c_s^2$, normalized by its value at $T_c$ in pure Yang-Mills theory (full circles) and QCD with $N_f=2+1$ (full squares).}
	\label{BulkDeltacs2}
\end{figure*}
%%%%%%%%%%%%%%%%%%%%%%%%%%%%%%%%%%%%%%%%%%%%%%%%%%%%%%%%%%%%%%%%%%%%%%%%%%%%%

%%%%%%%%%%%%%%%%%%%%%%%%%%%%%%%%%%%%%%%%%%%%%%%%%%%%%%%%%%%%%%%%%%%%%%
\subsection{Bulk to shear viscosity ratio\label{Subsec:BulkToShear}}
%%%%%%%%%%%%%%%%%%%%%%%%%%%%%%%%%%%%%%%%%%%%%%%%%%%%%%%%%%%%%%%%%%%%%%

Taking the high-temperature limit in Eqs.~(\ref{e:cs2}) and (\ref{zetaeq}), we find that the bulk viscosity $\zeta$ vanishes as the speed of
sound squared $c_s^2$ approaches the value $1/3$. Thus, the non-vanishing $\zeta$ near $T_c$ measures how far from the conformal limit the system is.
The bulk to shear viscosity ratio of an interacting photon gas~\cite{Weinberg:1971mx} and in scalar field theory~\cite{Jeon:1995zm}
is given unambiguously by
\begin{eqnarray}
\frac{\zeta}{\eta}= 15 \Big(\frac{1}{3}-c_s^2\Big)^2\,.
\label{pqcd1}
\end{eqnarray}
The shear and bulk viscosities have been evaluated at high temperature perturbatively~\cite{Arnold:2003zc,Arnold:2006fz}, from which
one finds that the ratio $\zeta/\eta$ follows quantitatively the same trend as in Eq.~(\ref{pqcd1}).

In contrast, for strongly-coupled theories along with \mbox{gauge$\slash$gravity} duality, the ratio behaves as~\cite{Buchel:2005cv}
\begin{eqnarray}
\frac{\zeta}{\eta}\propto \Big( \frac{1}{3} - c_s^2 \Big)\,.
\end{eqnarray}
Yet another non-perturbative approach, which describes the Yang-Mills plasma based on the Gribov-Zwanziger quantization, leads to the ratio 
$\zeta/\eta$ linearly proportional to the quantity $\Delta c_s^2 = 1/3 - c_s^2$~\cite{Florkowski:2015dmm}, and thus to an intriguing
agreement with the result from gauge-gravity duality.

In a similar QPM framework for pure Yang-Mills theory, it has been shown that the ratio $\zeta/\eta$ linearly depends on $\Delta c_s^2$
near the first-order phase transition temperature~$T_c$, whereas it scales quadratically with $\Delta c_s^2$ at high temperature~\cite{Bluhm:2009ef}.
To quantify the impact of the quasiquarks on the same quantity, we shall study the quark-flavor dependence encoded in the transport coefficients
within the QPM.
We recall that in the kinetic theory under the relaxation time approximation, the shear viscosity of the QGP for~$N_f=2+1$ \mbox{reads~\cite{Hosoya:1983xm,Gavin:1985ph,Sasaki:2008fg,Bluhm:2009ef,Chakraborty:2010fr,Dusling:2011fd}}
\begin{equation}
\label{eta}
\eta = \frac{1}{15 T}  \sum_{i=l,\bar{l}, s, \bar{s}, g} \int \frac{d^3p}{(2\pi)^3} \frac{p^4}{E_i^2}
d_i \tau_i f_i^0 (1 \pm f_i^0)\,.
\end{equation}
The shear viscosity to entropy density ratio in the two theories, pure Yang-Mills and QCD with $N_f=2+1$, has been explored in~\cite{Mykhaylova:2019wci},
and therein, the details on the $N_f$-dependence as well as a comprehensive comparison to other approaches are found.

The ratio of bulk to shear viscosity is readily calculated with Eqs.~(\ref{zetaeq}) and (\ref{eta}) in pure Yang-Mills and QCD with the corresponding effective masses.
Based on the observations in~\cite{Jaiswal:2016sfw, Bluhm:2009ef,Czajka:2018bod}, the full QPM results will be compared with the linear and
quadratic dependence on~$\Delta c_s^2$:
\begin{eqnarray}
\textrm{Linear:}\ \ \ \frac{\zeta}{\eta}= \alpha \Big( \frac{1}{3} - c_s^2\Big) + \beta, \label{approx_lin}\\
\textrm{Quadratic:}\ \ \ \frac{\zeta}{\eta}= \gamma \Big( \frac{1}{3} - c_s^2\Big)^2 + \delta, \label{approx_quad}
\end{eqnarray}
with fit parameters $\alpha,\beta,\gamma$ and $\delta$, under the assumption of equal relaxation times for the shear and bulk viscosities.
%%%%%%%%%%%%%%%%%%%%%%%%%%%%%%%%%%%%%%%%%%%%%%%%%%%%%%%%%%%%%%%%%%%%%%%%%%%%%%%%%%%%%%%%%%%
\begin{figure*}[htb!]
	\includegraphics[width=0.48\linewidth]{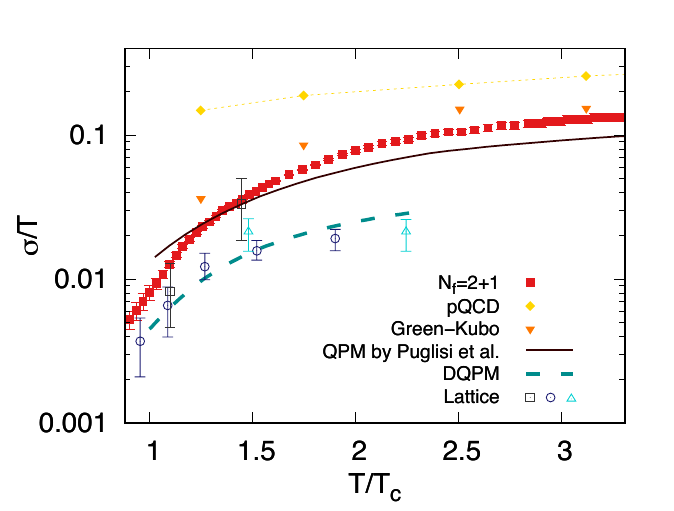}
	\includegraphics[width=0.48\linewidth]{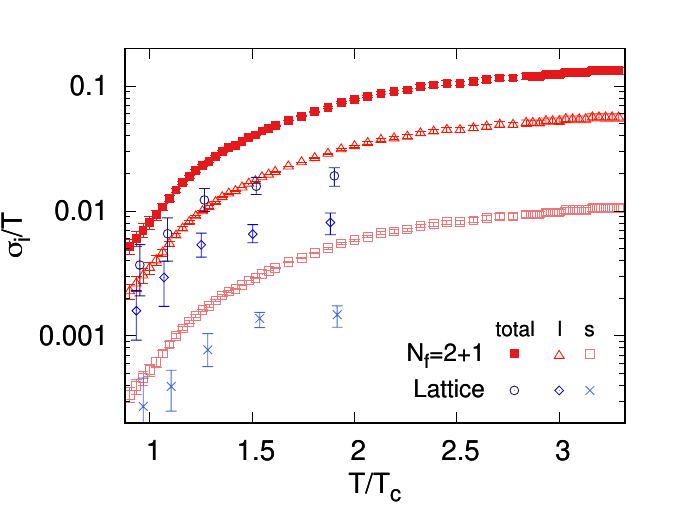}
	\caption{Left: Ratio of the total electrical conductivity to temperature as a function of $T/T_c$. Besides our QPM result (full squares), we show the results in other approaches collected in~\cite{Puglisi:2014sha}: the pQCD-based calculation (full diamonds), the Green-Kubo formalism (full triangles) and the QPM with a different setup (solid line). The dashed line corresponds to the DQPM result~\cite{Soloveva:2019xph}. The available results of lattice simulations are shown by open symbols: squares~\cite{Ding:2010ga}, triangles~\cite{Aarts:2007wj} and circles~\cite{Amato:2013naa}. Right: The total $\sigma/T$ ratio of the QGP (full squares) along with the light (open triangles) and strange (open squares) quark components.
The corresponding lattice data is deduced from~\cite{Aarts:2014nba} for the total electrical conductivity (open circles) and for light and strange quark contributions (diamonds and crosses, respectively). Here, our $\sigma_{l,s}/T$ ratios do not include the anti-particle contributions for a direct comparison to the lattice QCD data which are just for the particles.
		\label{fig:elcond}}
\end{figure*}
%%%%%%%%%%%%%%%%%%%%%%%%%%%%%%%%%%%%%%%%%%%%%%%%%%%%%%%%%%%%%%%%%%%%%%%%%%%%%%%%%%%%%%%%%
Figure~\ref{fig:BulkShearApprox} presents the bulk to shear viscosity ratio in pure Yang-Mills theory (left) and in QCD with \mbox{$N_f=2+1$ (right)}.
Consistently to the earlier study in Yang-Mills thermodynamics~\cite{Bluhm:2009ef}, the QPM result is well captured by the linear ansatz~(\ref{approx_lin}) near $T_c$, and by the quadratic one (\ref{approx_quad}) at higher temperature.
A clear changeover from the linear to quadratic scaling emerges at $T \simeq 1.3\,T_c$.
Near $T_c$ the $\zeta/\eta$ in the QPM agrees fairly well with the same quantity deduced from the available lattice data~\cite{Meyer:2007ic,Meyer:2007dy} and \cite{Astrakhantsev:2017nrs,Astrakhantsev:2018oue}.
Above $T \simeq 1.4\,T_c$, it is in line with the pQCD prediction~\cite{Arnold:2003zc,Arnold:2006fz}, wherein the shear and bulk viscosities
are given in the next-to-leading-log expansion by
\begin{eqnarray}
\eta_{\textrm{NLL}} = \frac{T^3}{g^4} \frac{\eta_1}{\ln(\mu_1^*/m_D)}\,,
\quad
\zeta_{\rm NLL} = \frac{A \alpha_s^2 T^3}{\ln(\mu_2^*/m_D)}\,,
\label{pQCDzeta}
\end{eqnarray}
with the strong coupling $\alpha_s=g^2/4\pi$ and the Debye mass squared \mbox{$m_D^2=(1+N_f/6)g^2 T^2$}.
For $N_f=0$, the set of parameters reads \mbox{$\eta_1=27.126$}, \mbox{$\mu_1^*/T=2.765$}, \mbox{$A=0.443$} and \mbox{$\mu_2^*/T=7.14$},
while for $N_f=3$, \mbox{$\eta_1=106.66$}, \mbox{$\mu_1^*/T=2.957$}, \mbox{$A=0.657$} and \mbox{$\mu_2^*/T=7.77$}.

The QPM with $N_f=2+1$ retains the same feature, but the changeover between the two scaling behaviors appears at a higher temperature, $T \simeq 2\,T_c$.
One observes a somewhat larger difference from the pQCD result, arising from the presence of quasiquarks. Including the matter fields also results in a delay of restoring conformal invariance at high temperature, as seen in the $\zeta/s$ ratio presented in Fig.~\ref{fig:CouplingDer+BulkSQPM} (right).

To look further into the relation between the bulk viscosity and conformality, in Fig.~\ref{BulkDeltacs2} we show a flavor dependence of the $\zeta/\eta$ ratio as a function of the measure $\Delta c_s^2$, as well as the explicit temperature profiles of $c_s^2$ at and above $T_c$.
One readily finds that (i) the speed of sound squared in QCD approaches its conformal value at high temperature, but much slower than in Yang-Mills
theory because of the presence of dynamical quarks, and (ii) the changeover of the two scaling behaviors in the $\zeta/\eta$ ratio is preserved.

%%%%%%%%%%%%%%%%%%%%%%%%%%%%%%%%%%%%%%%%%%%%%%%%%%%%%%%%%%%%%%%%%%%%%%%
\section{Electrical conductivity\label{Sec:Elcond}}

The electrical conductivity $\sigma$ quantifies the ability of a system to conduct the electric charge. In the relaxation time approximation,
the $\sigma$ of QGP with $N_f=2+1$ reads~\cite{Thakur:2017hfc}
\begin{eqnarray}
\sigma= \frac{1}{3 T} \sum_{i = u,\bar{u}, d, \bar{d},s,\bar{s}} \int \frac{d^3p}{(2 \pi)^3}  \frac{p^2}{E_i^2} q_i^2   d_i  \tau_i f_i^0 (1-f_i^0),\ \ \ 
\label{e:elcond}
\end{eqnarray}
where the quark electric charge $q_i$ is given explicitly by $q_u=-q_{\bar{u}}=2\,e/3$ and $ q_{d,\, s}=-q_{\bar{d},\, \bar{s}}=-e/3$.
The electron charge reads $e=(4 \pi \alpha)^{1/2}$ with the fine structure constant $\alpha\simeq 1/137$,
and the degeneracy factor in the above expression is $d_{u,\, d,\, s}=6$.
The contribution from light quarks will be denoted by $\sigma_l=\sigma_u + \sigma_d$.

In Fig.~\ref{fig:elcond} (left), we present the scaled electrical conductivity $\sigma/T$ including the results of various approaches. 
The QPM result is quite consistent with the earlier study~\cite{Puglisi:2014sha}, where $\sigma$ has been evaluated in the Green-Kubo formalism
and in the relaxation time approximation.
A~slight difference from the approach employed in~\cite{Puglisi:2014sha} arises from a few key features in modeling the QCD thermodynamics:
the effective coupling is parameterized as~\cite{Plumari:2011mk}
\begin{eqnarray}
g^2(T)= \frac{48 \pi^2}{(11 N_c-2 N_f)\ln\Big[\lambda(\frac{T}{T_c}-\frac{T_s}{T_c})\Big]^2},
\label{paramG}
\end{eqnarray}
with $\lambda=2.6$ and $T_s/T_c=0.57$ to reproduce the EoS in lattice QCD~\cite{Borsanyi:2013bia}.
Their quasiparticle masses are introduced as $m_g= 3 g^2 T^2/4$ and $m_q^2=g^2 T^2/3$, i.e., all quarks are degenerate.
The transport cross-sections used in~\cite{Puglisi:2014sha} depend on the Debye mass originated from the HTL approach,
\begin{eqnarray}
\sigma^{ij}_{tot}=\beta^{ij} \frac{\pi \alpha_s^2}{m_D^2} \frac{s}{s+m_D^2},
\label{pqcdcrosssec}
\end{eqnarray}
where $\beta^{ij}$ are the group factors responsible for different interactions between quarks and gluons: \mbox{$\beta^{qq}=16/9\,$}, $\beta^{qq'}=8/9,\ \beta^{qg}=2,\ \beta^{gg}=9$.
The result corresponding to perturbative QCD is deduced by setting the effective coupling to~\cite{Puglisi:2014sha}
\begin{eqnarray}
g_{\rm pQCD}= \frac{8 \pi}{9} \ln^{-1}\Big[\frac{2 \pi T}{\Lambda_{\rm QCD}}\Big].
\label{pqcdcoupling}
\end{eqnarray}

One finds an overall qualitative agreement with the DQPM result~\cite{Soloveva:2019xph}, although their approach yields a somewhat smaller $\sigma/T$
at any temperature. A similar trend was already observed in the $\zeta/s$ ratio (Fig.~\ref{fig:CouplingDer+BulkSQPM}-right).
Furthermore, as briefly discussed in Sec.~\ref{Subsec:BulkToEntropy}, the large angle scattering approximation in evaluating the cross-section 
leads to a systematic upward-shift of any transport parameters.

The QPM also well captures the behavior in the vicinity of $T_c$, consistently to that found in the lattice calculations~\cite{Ding:2010ga,Aarts:2007wj,Amato:2013naa}.
To look into the role of different quark flavors, we present the contributions from light and strange quarks along with the total electrical conductivity
in Fig.~\ref{fig:elcond} (right).
The light-quark contribution is larger than that of the strange quarks, as anticipated with their mass differences shown in Fig.~\ref{fig:massestot}.
The corresponding lattice data~\cite{Aarts:2014nba} near $T_c$ is rather compatible to the QPM result, whereas the discrepancy between them emerges
at $T \simeq 1.5\, T_c$ and increases gently with temperature.
This can be attributed to the fact that the lattice setup includes the pion mass $M_\pi = 384(4)$ MeV, heavier than the physical one
used in our model-building.
In fact, by increasing the bare mass of light quarks, we obtain a decrease of the electrical conductivity, and this is smaller than the result with physical quark masses at any temperature.

\section{Summary\label{Sec:Conclusions}}
%%%%%%%%%%%%%%%%%%%%%%%%%%%%%%%%%%%%%%%%%%%%%%%%%%%%%%%%%%%%%%%%%%%%%%%%%%%%%%%%%%%%%%%%%%%

We have examined the transport coefficients, \mbox{the bulk~$\zeta$} and shear $\eta$ viscosity, and the electrical conductivity $\sigma$, of deconfined
strongly interacting matter in pure Yang-Mills theory and QCD with $N_f=2+1$ at vanishing chemical potential. We employed the kinetic approach
for a medium whose thermodynamics is described in a quasiparticle model (QPM) under the relaxation time approximation.
The dynamical masses of quasiparticles are characterized by an effective running coupling depending explicitly on temperature, deduced from
the entropy density in lattice simulations with different quark flavors, $N_f=0$ and $2+1$.
To verify the validity of the QPM to bulk thermodynamic quantities near the phase transition, we computed the speed of sound squared, $c_s^2$,
in the QPM and the hadron resonance gas model, and confronted it with the corresponding lattice results.
It is found that the QPM captures extremely well not only the behavior at high temperature but also that in the vicinity of the phase transition
and even slightly below~$T_c$. The~$c_s^2$ below but near $T_c$ requires a tower of hadronic resonances, and this non-trivial physics is
properly encoded in the effective coupling.

Assuming that all the transport coefficients studied in this paper are characterized by the same relaxation times~$\tau_i$, we used the total cross
sections calculated for the elementary two-body scattering processes of the quasiquarks and gluons given in~\cite{Mykhaylova:2019wci}.
In pure Yang-Mills theory, the temperature derivative of the gluon effective mass yields a striking peak at the critical temperature, and this,
though much weakened, results in a mild non-monotonicity in the bulk viscosity to entropy density ratio~$\zeta/s$.
The bulk viscosity decreases as temperature increases, consistently to the general anticipation, and conformal invariance becomes restored
at high temperature.
Including light and strange quasiquarks considerably modifies  the $\tau_i$ and $\zeta$ as well as the entropy density~$s$.
For the QGP with $N_f=2+1$, the ratio $\zeta/s$ does not exhibit any apparent non-monotonicity around the crossover, and decreases with increasing temperature
much slower than in the $N_f=0$ case, indicating a larger breaking of a scale symmetry.

Given the bulk and shear viscosities, we constructed the ratio $\zeta/\eta$ to confront with the linear and quadratic dependence on the measure
$\Delta c_s^2 = 1/3 - c_s^2$ representing a deviation from conformal invariance.
We find that the ratio scales linearly near $T_c$, as predicted in the AdS/CFT approach~\cite{Buchel:2005cv}, then switches to the quadratic behavior
consistently to the perturbative QCD result~\cite{Arnold:2003zc, Arnold:2006fz}. The emerging changeover depends on the quark flavors: in pure Yang-Mills theory
it appears at $T \simeq 1.3\, T_c$, whereas in QCD with $N_f=2+1$ at $T \simeq 2\, T_c$.
Thus, the segment in temperature where one finds the system non-perturbative is interestingly extended in the presence of dynamical quasiquarks.
The QPM well captures the smooth but $N_f$-depending changeover to describe the non-perturbative and perturbative domains.
We also found that the presence of quasiquarks results in a significant delay of restoring conformal invariance at high temperature, compared with
pure Yang-Mills thermodynamics.

We also studied the electrical conductivity of the QGP with $N_f=2+1$. The ratio $\sigma/T$ is found to be qualitatively consistent with earlier
results in a class of QPM~\cite{Puglisi:2014sha,Soloveva:2019xph} as well as with the recent lattice QCD results~\cite{Aarts:2014nba}.
In particular, the individual contributions to the electrical conductivity were calculated separately for the light and strange quarks, and
confronted with the corresponding lattice data. We find that the behaviors are even quantitatively close to the lattice results but systematically
lower than those. This is explained by the fact that the simulations in \cite{Aarts:2014nba} carried out for a heavy pion mass, $M_\pi \approx 384$ MeV.

We have quantified the impact of dynamical quarks on the major transport parameters. The QPM has the capability to describe systematically
a non-trivial link between the non-perturbative and perturbative physics relevant for the transport properties of the deconfined matter.
As explored recently in~\cite{HJG}, the resultant transport coefficients can be quantitatively very different depending on the prescriptions
how to deal with the non-equilibrium nature. Besides, a more realistic estimate may require further extensions going beyond the major assumptions made in this paper,
i.e., the momentum-independent relaxation times common for the shear and bulk viscosities, and the electrical conductivity.
Those can be implemented into our kinetic approach as guided in \cite{Jeon:1995zm,Czajka:2017bod,Chakraborty:2010fr,HJG} in offering more
reliable medium-profiles of the transport coefficients for hydrodynamic simulations.

%%%%%%%%%%%%%%%%%%%%%%%%%%%%%%%%%%%%%%%%%%%%%%%%%%%%%%%%%%%%%%%%%%%%%%%%%%%%%%%%%%%%%%%%%%%
\acknowledgments
The authors thank Marcus Bluhm, Pok Man Lo, Micha{\l} Marczenko and Krzysztof Redlich for stimulating discussions.
This work was partly supported by the Polish National Science Center (NCN) under the Opus grant no. 2018/31/B/ST2/01663.
%%%%%%%%%%%%%%%%%%%%%%%%%%%%%%%%%%%%%%%%%%%%%%%%%%%%%%%%%%%%%%%%%%%%%%%%%%%%%%%%%%%%%%%%%%%
\appendix
\section{Energy-dependent relaxation time}
\label{sec:appendix}
\begin{figure}[htb!]
	\includegraphics[width=1\linewidth]{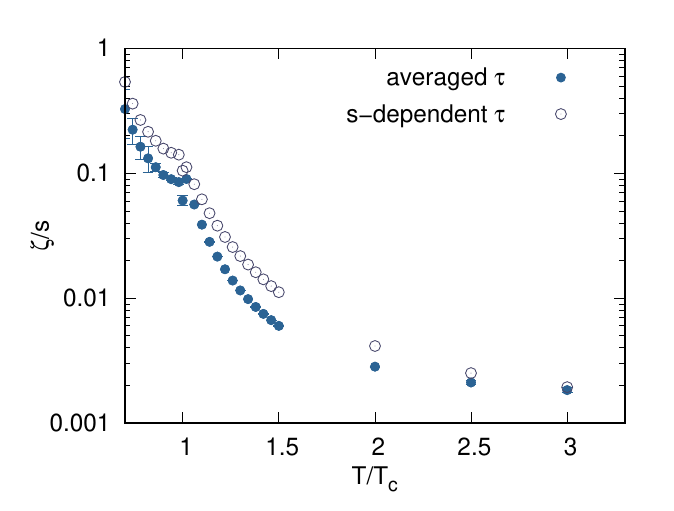}
	\caption{The bulk viscosity to entropy density ratio as a function of $T/T_c$ in pure Yang-Mills theory. The result obtained with the mean relaxation time is shown by the full circles, while the open bullets represents the $\zeta/s$ ratio based on the energy-dependent $\tau(s)$.}
		\label{fig:zeta_appendix}
\end{figure}

All transport parameters in the QPM are computed with the mean relaxation time based on the thermal-averaged total cross sections of the quasiparticles. To illustrate how it works, we evaluate the specific bulk viscosity in pure Yang-Mills theory using the energy-dependent relaxation time as well in this Appendix. A~comparison to the $\zeta/s$ ratio with the mean~$\tau$~is~shown~in~Fig.~\ref{fig:zeta_appendix}.

It is clear that both results share the same qualitative trend and that they are even quantitatively close to each other. 
This feature is also expected in QCD with $2+1$ quark flavors since there exists just a smooth crossover, not anything drastic like a phase transition.
Thus, in view of the main goal of our study, it is sufficient to use the energy-averaged relaxation time to capture the correct physics and to clarify the role of dynamical quarks in the transport properties of the QGP.
%%%%%%%%%%%%%%%%%%%%%%%%%%%%%%%%%%%%%%%%%%%%%%%%%%%%%%%%%%%%%%%%%%%%%%%%%%%%%%%%%%%%%%%%%%%

\end{document}